\documentclass[letterpaper, 10 pt, conference]{ieeeconf}  

\usepackage{cite}
\usepackage{amsmath,amssymb,amsfonts}
\usepackage{graphicx}
\usepackage[colorlinks]{hyperref}
\hypersetup{citecolor=[RGB]{5,112,176},linkcolor=[RGB]{5,112,176},urlcolor=[RGB]{5,112,176}}

\usepackage{algorithm}
\usepackage{algpseudocode}
\usepackage{tikz}

\usepackage{textcomp}

\usepackage{mathtools}
\usepackage{bm}
\usepackage{bbm}

\usepackage{graphicx}
\usepackage{amssymb}

\usepackage{algorithm}
\usepackage{algpseudocode}

\newcommand{\R}{\mathbb{R}} 
\newcommand{\CR}{\mathcal{R}} %
\newcommand{\CC}{\mathcal{C}} %
\newcommand{\CE}{\mathcal{E}} %
\newcommand{\CM}{\mathcal{M}} %
\newcommand{\col}{{\rm col}} %

\usepackage{amsthm}

\newtheorem{theorem}{Theorem}

\newtheorem{lemma}{Lemma}

\theoremstyle{definition}

\newtheorem{example}{Example}

\newcommand*{\QE}{\hfill\ensuremath{\blacksquare}}	
\newcommand{\TiaQED}{\hfill $\triangle$} 			

\newcommand{\references}{refers.bib}

\IEEEoverridecommandlockouts                              

\title{\LARGE \bf
Modeling and Detection of Critical Slowing Down in Epileptic Dynamics
}

\author{Yuzhen Qin, Marcel van Gerven
\thanks{This work was supported in part by the project Dutch Brain Interface Initiative (DBI$^2$) with project number 024.005.022 of the research programme Gravitation which is (partly) financed by the Dutch Research Council (NWO). Y. Qin and M. van Gerven are with the Department of Machine Learning and Neural Computing, Donders Institute for Brain, Cognition and Behaviour, Radboud University, Nijmegen, the Netherlands. 
        {\tt\small \{yuzhen.qin, marcel.vangerven\}@donders.ru.nl}}%
}

\begin{document}

\maketitle
\thispagestyle{empty}
\pagestyle{empty}

\begin{abstract}
	Epilepsy is a common neurological disorder characterized by abrupt seizures. Although seizures may appear random, they are often preceded by early warning signs in neural signals, notably, critical slowing down, a phenomenon in which the system’s recovery rate from perturbations declines when it approaches a critical point. Detecting these markers could enable preventive therapies. This paper introduces a multi-stable slow-fast system to capture critical slowing down in epileptic dynamics. We construct regions of attraction for stable states, shedding light on how dynamic bifurcations drive pathological oscillations. We derive the recovery rate after perturbations to formalize critical slowing down. A novel algorithm for detecting precursors to ictal transitions is presented, along with a proof-of-concept event-based feedback control strategy to prevent impending pathological oscillations. Numerical studies are conducted to validate our theoretical findings.
\end{abstract}

\section{INTRODUCTION}
Epilepsy is one of the most common and serious chronic neurological disorders \cite{2020_LW_PH_SSM_el_al}, characterized by the sudden onset of harmful seizures. Current interventions—such as medication, surgery, and neuromodulation—offer limited effectiveness due to their empirical nature~\cite{meisel2020seizure}. This highlights an urgent need for systematic, mechanistic approaches to develop optimized, individualized therapies. Modeling epilepsy dynamics offers a promising approach to understanding individual mechanisms, fostering the development and testing of personalized, optimized therapies. 

Epilepsy involves two primary brain states: the \textit{interictal} state, characterized by normal neural activity, and the \textit{ictal} one, marked by excessively synchronized oscillations~\cite{2003_DSFGP_BW_et_al}.
A variety of models use bistability to capture these two states \cite{2010_FF_STJ_BM}, with a stable equilibrium representing normal neural activity and a stable limit cycle representing pathological oscillations. Transitions between these states model seizure onset and offset. Studies suggest that transitions can be triggered by internal disturbances or external stimuli \cite{2012_BO_FTHB_Math_epilepsy}, aligning with the dynamics observed in absence seizures \cite{snead1995basic}. 

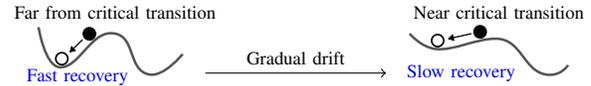
\begin{figure}[t]
	\centering
	\begin{tikzpicture}
		\node at(-1,0.1) {\includegraphics[scale=0.6]{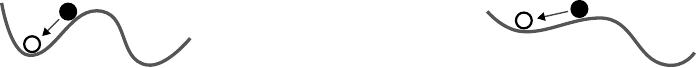}};		
		\node[scale=0.7] at(-3.5,0.6) {Far from critical transition};
		\node[scale=0.7] at(1.6,0.6) {Near  critical transition};		
		\node[scale=0.7,blue] at(-4,-0.25) {Fast  recovery};
		\node[scale=0.7,blue] at(1.1,-0.2) {Slow  recovery};
		\node[scale=0.7] at(-1.1,0) {Gradual drift};
		\draw[->] (-2.3,-0.2) to (0.1,-0.2); 
	\end{tikzpicture}
	\vspace{-10pt}
	\caption{Illustration of  critical slowing down }		
	\label{cricical}
	\vspace{-20pt}
\end{figure}

However, simple bistable mechanisms may not fully capture the dynamics of all epilepsy types. In many cases, such as temporal lobe epilepsy, seizures arise from gradual changes at cellular, molecular, and network levels in the brain \cite{tatum2012mesial}. Some studies suggest that these transitions are tied to the brain's gradual drift toward a critical threshold, progressively destabilizing healthy states. Upon crossing the critical point, such normal states transition to pathological ones abruptly \cite{chang2018loss}. Interestingly, neural signals at the verge of the critical point often exhibit \textit{critical slowing down}  \cite{scheffer2009early}, a phenomenon characterized by slower recovery times after perturbations (see Fig.~\ref{cricical}). This has been explored as a potential biomarker for seizure prediction \cite{maturana2020critical}. A rigorous understanding of critical slowing down holds significant potential for designing effective schemes to detect early warning signs of seizures and for developing timely preventive interventions. 


\textbf{Related work.}  Researchers have employed networks of linear-threshold units to study epileptic dynamics, focusing on bifurcations between equilibria and limit cycles  \cite{2022_NE_PR_CJ:Auto,2023_MM_MT_CJ:LCSS,2021_CF_AA_PF_CJ:LCSS}. Additionally, several studies have developed network control strategies aimed at preventing the spread of pathological oscillations \cite{2022_AA_CF_PF_CJ:OJCS}  or restoring normal states \cite{2022_own_CDC_a,2023_own_OJCS,2023_own_ACC}. Other approaches use linear fractional-order systems \cite{2023_REA_RG_BP_PS_ACC, chatterjee2020fractional} and bilinear systems\cite{acharya2024predictive} to model and regulate epilepsy. However, none of these works address the modeling and detection of critical slowing down. Recent work offers methods to detect critical slowing down in network systems \cite{acharya2024predictive}, but they rely on passive monitoring of the states and work only when states remain close to the equilibrium.

\textbf{Contribution.} This paper provides an analytical exploration of critical slowing down in epileptic dynamics with four main contributions. First, we introduce a multi-stable slow-fast system to model epileptic dynamics, where a slow state captures changes in neuronal excitability, leading to dynamic bifurcations that initiate or terminate epileptiform oscillations by shifting between stable states. Second, we construct regions of attraction for these states, offering a theoretical basis for transitions between normal neural activity and pathological oscillations. Third, we derive a convergence rate for any perturbations within these attraction regions, providing formal evidence of critical slowing down. This is different from existing approaches that rely on linearization, which only applies to small, near-equilibrium perturbations \cite{scheffer2009early, acharya2024predictive}.
Fourth, leveraging this derived convergence rate, we design an active detection scheme that probes and monitors the system to identify pre-ictal events, setting it apart from traditional methods relying on passive recordings \cite{chang2018loss, maturana2020critical}. Additionally, we present a proof-of-concept feedback control strategy that prevents seizure onset by responding to detected precursors, laying a foundation for preventive therapies for epilepsy with minimal intervention. Our theoretical results are demonstrated by numerical studies.



\section{Problem Formulation}\label{prom_formu}
In our recent work \cite{self_qin2024analytical}, we model the dynamics of neuronal populations using the following system:
\begin{subequations}\label{old}
    \begin{align}
    \dot x &= - \omega y + x f(x, y) , \\
    \dot y &= \hphantom{-}\omega x + y f(x, y),
\end{align}
\end{subequations}
where $f(x,y) \coloneq \sigma + 2 a b (x^2+y^2)-b(x^2+y^2)^2$, and $x,y\in \R$ represent the population-averaged firing rate of pyramidal and interneuron ensembles, respectively \cite{2014_KS_KM_PG_DA_FL}. The parameters are constrained to $\omega, a, b>0$, and $\sigma\in \R$. 

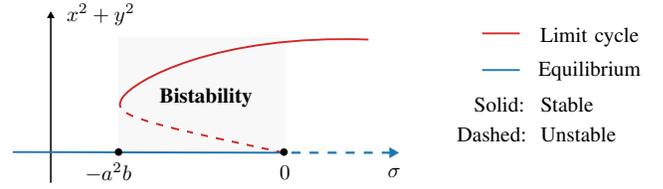
\begin{figure}[t]
	\centering
	\begin{tikzpicture}
		\node at(-1,0.1) {\includegraphics[scale=1]{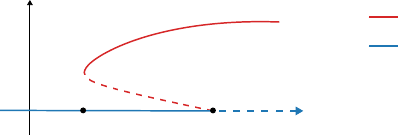}};
		
		\node[scale=0.8] at(-3.2,1.2) {$x^2+y^2$};
		\node[scale=0.8] at(0.7,-.9) {$\sigma$};
		\node[scale=0.8] at(-3.1,-.9) {$-a^2b$};
		\node[scale=0.8] at(-0.75,-.9) {$0$};
		
		\node[scale=0.8] at(3.3,0.9) {Limit cycle};
		\node[scale=0.8] at(3.3,.45) {Equilibrium};
		\node[scale=0.8] at(2.1,0) {Solid: };
		\node[scale=0.8] at(3,0) {Stable};
		
		\node[scale=0.8] at(-1.8,.1) {\textbf{Bistability}};
		\node[scale=0.8] at(2,-0.4) {Dashed: };
		\node[scale=0.8] at(3.15,-.4) {Unstable};
		
		\fill[opacity=0.03] (-2.96,-0.65) rectangle (-0.75,0.9);		
	\end{tikzpicture}
	\hspace{-10pt}
	\caption{Bifurcation diagram of system \eqref{old} illustrating the emergence and disappearance of equilibria and limit cycles across parameter regimes.}		
	\label{bifurcation}
\end{figure}

Figure~\ref{bifurcation} illustrates diverse behaviors of the system across different parameter regimes. When $\sigma<-a^2 b$, the equilibrium $\col(x,y) \coloneq [x,y]^\top=0$ is globally exponentially stable. At $\sigma=-a^2 b$, a saddle-node bifurcation occurs, giving rise to two limit cycles when $\sigma>-a^2 b$:
\begin{align*}
	&\CC \coloneq\{x,y\in\R\colon x^2 +y^2 = {a + \sqrt{a^2 + \sigma/b}}\},\\
	&\CC' \coloneq\{x,y\in\R\colon x^2 +y^2 = {a - \sqrt{a^2 + \sigma/b}}\}.
\end{align*}
In this regime, both the equilibrium $\col(x,y)=0$ and the limit cycle $\CC$ are stable, with their regions of attraction separated by the unstable limit cycle $\CC'$.  At $\sigma=-a^2 b$, the system undergoes a saddle-node bifurcation of limit cycles. Here, the unstable limit cycle $\CC'$ collides with the equilibrium, destabilizing it and leaving the limit cycle $\CC$ as the unique stable attractor for $\sigma>0$. 

In the bistable regime highlighted by the shaded gray area in Fig.~\ref{bifurcation}, the system \eqref{old} effectively models core dynamics seen in certain epilepsy types associated with absence seizures \cite{snead1995basic}. Here, the equilibrium represents stable, normal brain activity, while the stable limit cycle reflects seizure-like oscillations. Seizure onset and offset are driven by random perturbations, mimicking the spontaneous nature of these episodes. As we detailed in \cite{self_qin2024analytical}, the region of attraction surrounding the equilibrium serves as a key feature,  distinguishing healthy brains from epileptic ones. A reduced region of attraction implies greater sensitivity to disturbances---a characteristic often associated with epilepsy. The parameter $\sigma$ plays a crucial role in modulating the size of this attraction region, thereby influencing neuronal excitability, with larger $\sigma$ corresponding to increased excitability.

Absence seizures are challenging to predict due to their spontaneous nature and sensitivity to random, unpredictable perturbations \cite{kuhlmann2018seizure}. In contrast, seizures arising from temporal lobe epilepsy and similar conditions often stem from gradual, progressive changes at cellular, molecular, and network levels in the brain  \cite{tatum2012mesial}. These gradual changes often have slow timescales ranging from hours and days to even months \cite{baud2018multi, karoly2021cycles}, significantly slower than typical neural activity. Upon reaching a critical threshold, they abruptly transform healthy states into pathological ones. Near this critical point, neural activity often demonstrates ``critical slowing down,'' where recovery from perturbations slows as the system approaches instability. This property can be leveraged to predict seizures and design preventive therapies. However, system \eqref{old} lacks the capability to capture this aspect of neural dynamics, limiting its effectiveness for forecasting such critical transitions.

This paper seeks to extend the model in \eqref{old} to achieve three main goals: (i) capture a wider range of epilepsy dynamics; (ii) provide a theoretical foundation for understanding critical slowing down in relation to seizure onset; and (iii) leverage these insights to design strategies for early detection of seizure warning signs and timely intervention.

\section{Modeling via slow-fast systems}
One observes that in \eqref{old}, the excitability parameter $\sigma$ is fixed. However, due to the non-static nature of the brain, a more realistic assumption is that $\sigma$ is dynamically changing. Motivated by this, we propose a slow-fast system to describe epileptic dynamics:
\vspace{-6pt}
\begin{subequations}\label{main}
	\begin{align}
		\dot x &= - \omega y + x f(x, y, \sigma),\label{main_fast1} \\
		\dot y &= \hphantom{-}\omega x + y f(x, y,\sigma),\label{main_fast2}\\
		\dot \sigma & = -\varepsilon (\sigma-c_1)(\sigma-c_2)(\sigma- c_3), \label{main_slow}
	\end{align}
	where $
		f(x,y,\sigma) \coloneq \sigma + 2 a b (x^2+y^2)-b(x^2+y^2)^2. 
	$
\end{subequations}
The function $g$ governs the dynamics of the excitability parameter. In real-world scenarios, the dynamics of $\sigma$ could also depend on the neural activity $x,y$ and internal perturbations or external stimuli. However, we restrict our attention to this simple $g$ for simplicity, which, as it turns out later, is sufficient to capture a wide range of epileptic dynamics. The parameters $c_1,c_2,c_3 \in \R$ satisfy $c_1<c_2<c_3$.
The constant $\varepsilon>0$ is typically very small, indicating that changes in the excitability of neuronal populations occur on a much slower timescale than the neural activity itself.

\subsection{Multistability}

Notice that the slow subsystem \eqref{main_slow} has three equilibria, $\sigma=c_1,c_2,c_3$. For each of them, we associate an equilibrium and two manifolds of the overall system \eqref{main}, respectively:
\begin{subequations}\label{manifolds}
    \begin{align*}
    &\CE_i \coloneq \{x,y,\sigma\in\R\colon x=0,y=0,\sigma=c_i\},\\
    &\CM_i \coloneq \{x,y,\sigma\in\R\colon x^2+y^2=a+\gamma_{i},\sigma=c_i\},\\
    &\CM_i' \coloneq \{x,y,\sigma\in\R\colon x^2+y^2=a-\gamma_{i},\sigma=c_i\},
\end{align*}
\end{subequations}
where 
$
    \gamma_{i} \coloneq \sqrt{a^2+c_i/b},
$
for all $i=1, 2, 3$. The equilibria describe normal neural activity and the manifolds, $\CM_i$'s in particular, capture pathological oscillations. 

It can be observed that the equilibria always exist, but whether a manifold exists depends on the parameters. It follows from Section~\ref{prom_formu} that  $\CM_i$ exists only when $c_i\ge-a^2 b$, while $\CM_i'$ exists only when $-a^2 b< c_i<0$. 

For the slow subsystem \eqref{main_slow}, one can derive that $\sigma=c_1$ and $\sigma=c_3$ are stable, and $\sigma=c_2$ is unstable. Therefore, $\CE_2, \CM_2$, and $\CM_2'$, when they exist, are always unstable. The stability of $\CE_1, \CE_3, \CM_1, \CM_1', \CM_3$, and $\CM_3'$ depends on the parameters, which are characterized by the lemma below.
\begin{lemma}
    For the system \eqref{main}, the following statements hold for $i=1, 3$:
    \begin{enumerate}
        \item[(i)] if $c_i< -a^2b$, $\CE_i$ is exponentially stable, while $\CM_i$ and $\CM_i'$ do not exist.
        \item[(ii)] if  $-a^2b<c_i<0$, $\CE_i$ and $\CM_i$ are exponentially stable, while $\CM_i'$ is unstable.
        \item[(ii)] if $c_i>0$, $\CM_i$ is exponentially stable, while $\CE_i$ is unstable and $\CM_i'$ does not exist.
    \end{enumerate}
\end{lemma}

For different configurations of parameters, the system can exhibit a wide range of multistable behaviors. In what follows, we focus on the situation where $-a^2b<c_1<c_2<0<c_3$, and show how the system \eqref{main} in this regime captures the dynamics of different epilepsy types.

\subsection{Modeling epileptic dynamics}
When $-a^2b<c_1<c_2<0<c_3$, the system has three attractors, $\CE_1, \CM_1$, and $\CM_3$. It can be observed that for any initial condition taken from 
\begin{equation*}
    \{x,y,\sigma \in \R: \col(x,y)\neq 0,\sigma>c_2\},
\end{equation*}
the solution to the system \eqref{main} converges to $\CM_3$. However, starting from $\sigma(0)<c_2$, the system exhibits bistability. The following theorem presents the region of attraction for the equilibrium $\CE_1$, demonstrating conditions under which the system can transition between $\CE_1$ and $\CM_1$.

\begin{theorem}
    Assume that the parameters in system \eqref{main} satisfy $-a^2b<c_1<c_2<0<c_2$. Then, both of the sets,
    \begin{align*}
       & \CR_1:=\{x,y,\sigma\in \R:\\
       &\hphantom{\CR_1:=\{x,y}x^2+y^2<a-\sqrt{a^2+\sigma/b},c_1\le\sigma<c_2\},\\
       & \CR_2:=\{x,y,\sigma\in \R: \sigma<c_1\},
    \end{align*}
    are positively invariant. In addition, for any initial condition that satisfies $\col(x(0),y(0),\sigma(0))\in \CR\coloneq \CR_1\cup \CR_2$, the solution $\col(x(t),y(t),\sigma(t))$ converges to $\CE_1$ exponentially. 
\end{theorem}

\begin{proof}  
	We first show $\CR_1$ is positively invariant and starting within $\CR_1$, the system converges to $\CE_1$ exponentially. 
    For any initial condition in $\CR$, $\sigma(0)<c_2$. Then, it can be
    derived that the solution $\sigma(t)$ of the subsystem \eqref{main_slow} converges to $\sigma=c_1$ exponentially. In addition, we have $\sigma(t)\le 
    \sigma(0)$ for any initial condition $\sigma(0)<c_2$, implying that $\{\sigma\in R:c_1\le \sigma<c_2\}$ is positively invariant. 

    For the fast subsystems \eqref{main_fast1} and \eqref{main_fast2}, consider the Lyapunov function candidate $V(x,y)=\frac{1}{2}(x^2+y^2)$. Its time derivative satisfies
   $
        \dot V(x,y) = (x^2+y^2)f(x,y,\sigma(t)).
    $
    Since $\sigma(t)\le \sigma(0)<c_2$ for all $t\ge 0$, we have 
    \begin{equation*}
        \dot V(x,y) \le(x^2+y^2)f(x,y,\sigma(0)),
    \end{equation*}
    where the property that $f(x,y,\sigma)$ is increasing in $\sigma$ has been used.     
    For any $\col(x,y)\in \{x,y\in \R: x^2+y^2<a-\sqrt{a^2+\sigma(0)/b}\}\coloneq \CR_{\sigma(0)}$, it can be derived that 
    $
        \dot V(x,y)\le 0,
    $
    which implies that $\CR_{\sigma(0)}$ is positively invariant. Furthermore, one can see that $\dot V(x,y)\le -\alpha V(x,y)$ for some positive $\alpha$, indicating that the solution of \eqref{main_fast1} and \eqref{main_fast2} converges to $\col(x,y)=0$ exponentially. 

    From the above analysis, one can observe that starting from any $\sigma(0)$ that satisfies $c_1\le \sigma(0)<c_2$, and $\col(x,y) \in \CR_{\sigma(0)}$, the solution $\col(x(t),y(t),\sigma(t))$ remains in the set $\{x,y,\sigma\in \R: x^2+y^2<a-\sqrt{a^2+\sigma(0)/b},  c_1\le \sigma(t)\le \sigma(0)\}$. In addition, $\col(x(t),y(t),\sigma(t))$ converges to $(0,0,c_1)$ exponentially. 
    
    Following similar steps, one can show that $\CR_2$ is also positively invariant and starting within $\CR_2$, the system converges to  $\CE_1$ exponentially, too.     The proof is complete.
    \end{proof}
    
    \begin{figure}[t]
    	\centering
    	\begin{tikzpicture}
    		\node at(-1,0.1) {\includegraphics[scale=1]{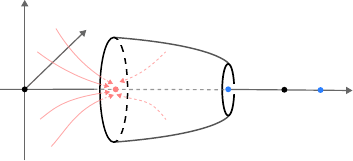}};
    		
    		\node[scale=0.8] at(-3.3,1.3) {$y$};
    		\node[scale=0.8] at(-2.5,1.1) {$x$};
    		\node[scale=0.8] at(2,-.3) {$\sigma$};
    		\node[scale=0.8] at(-4,-.3) {$-a^2b$};
    		\node[scale=0.8] at(0.81,-.3) {$0$};
    		
    		\node[scale=0.8] at(1.43,-.3) {$c_3$};
    		
    		\node[scale=0.8] at(-0.1,-.8) {$c_2$};
    		
    		\node[scale=0.8] at(-2,-.3) {$c_1$};
    	\end{tikzpicture}
    	\vspace{-8pt}
    	\caption{Region of attraction of steady state $\CE_1$. If $\sigma(0)<c_1$, the system will always converge to $\CE_1$ for any $x(0)$ and $y(0)$. For $c_1<\sigma(0)<c_2$, the convergence occurs within the cylindrical region. When $\sigma(0)>c_2$, the system will always converge to undesired oscillations. }		
    	\label{region_attrac}
    \end{figure}
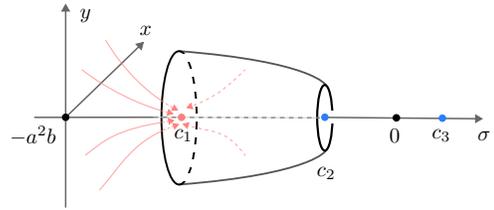


The system \eqref{main} in this regime is well-suited to model epileptic dynamics. To illustrate this point, we consider a forced version of it, resulting in
\vspace{-6pt}
\begin{subequations}\label{stochastic}
	\begin{align}
		\dot x &= - \omega y + x f(x, y, \sigma)+\zeta_x(t),\label{stochastic_fast1} \\
		\dot y &= \hphantom{-}\omega x + y f(x, y,\sigma)+\zeta_y(t),\label{main_slow_fast2}\\
		\dot \sigma & = -\varepsilon (\sigma-c_1)(\sigma-c_2)(\sigma- c_3)+\zeta_\sigma(t), \label{stochastic_slow}
	\end{align}
\end{subequations}
where the inputs $\zeta_x, \zeta_y,$ and $\zeta_\sigma$ can describe endogenous noise, exogenous disturbances, or external stimuli. 

Normally, the system operates near the equilibrium $\CE_1$, reflecting steady-state neural activity. Perturbations, described by $\zeta_x,\zeta_y,$ and $\zeta_\sigma$, can drive the system into pathological oscillations described by $\CM_1$ and $\CM_3$. When perturbations are sufficiently large to push the states outside $\CR$ simply along the $x$ and $y$ directions, transitions to ictal states occur abruptly. This situation captures the dynamics of absence seizures.  By contrast, when perturbations shift $\sigma$ just above the threshold $c_2$ ---the separatrix between the two stable states, $\sigma = c_1$ and $\sigma = c_3$, in the subsystem \eqref{main_slow} ---the transition to pathological oscillations can develop gradually over an extended period. 
 We illustrate this scenario with an example below.

\begin{figure}[t]
	\centering
	\includegraphics[scale=0.75]{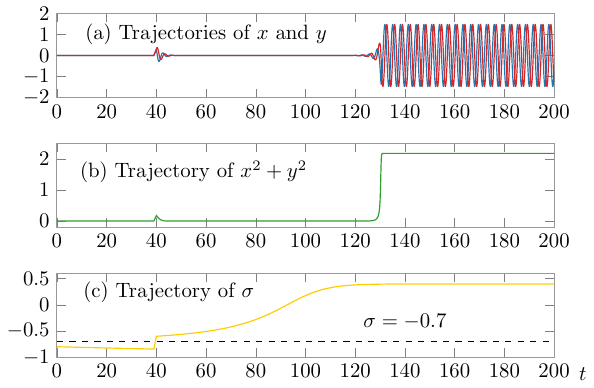}
	\caption{Perturbations at $t=40$ drive $\sigma$ beyond $-0.7$. Despite initial convergence towards the steady state, $x$ and $y$ quickly transition to pathological oscillations after $\sigma$ passes the critical point $0$. Parameters: $a=b=1, \omega=2, c_1 =-0.9, c_2=-0.7, c_3=0.5$, and $\varepsilon=0.1$.}	
	\label{pertub_to_oscillation}
 \vspace{-15pt}
\end{figure}

\begin{example}
Consider the system with parameters given in Fig.~\ref{pertub_to_oscillation}. At $t=40$, we introduce perturbations to $x$, $y$, and $z$. These adjustments are carefully designed so that they do not drive the states outside the region $\CR$ along the $x$ and $y$ directions, but shift $\sigma$ just above $c_2$. Initially, these perturbation-induced oscillations appear to be normal, decaying back toward equilibrium. However, because $\sigma$ has shifted beyond the equilibrium region of attraction for $\sigma=c_1$, it begins to slowly drift towards $\sigma=c_3$. Upon reaching $\sigma=0$, a saddle-node bifurcation destabilizes the equilibrium $\col(x,y)=0$. Consequently, the neural activity rapidly transitions to seizure-like oscillations as the stable equilibrium vanishes. \TiaQED
\end{example}

This example demonstrates that, besides noise-driven transitions,  our model can also capture transitions induced by \textit{dynamic bifurcations} that unfold more gradually. This slower, bifurcation-driven progression mirrors the gradual changes leading to seizure onset in epilepsy types such as temporal lobe epilepsy \cite{tatum2012mesial}. Observations of critical slowing down---a phenomenon where the recovery rate from small perturbations decreases---have been reported in neural recordings before seizure onset. This raises key questions: (1) Does the slow-fast system in \eqref{main} display critical slowing down as it nears the transition point? (2) Can critical slowing down be detected as an early warning signal for impending seizures?

\section{Detection of critical slowing down}
To address these questions, we begin by examining the response of system \eqref{main} to perturbations across various parameter regimes, as illustrated in the following example.

\begin{figure}[t]
	\centering
	\includegraphics[scale=0.75]{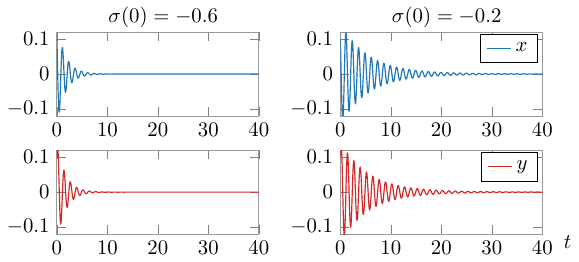}
	\caption{Comparison of convergence speeds for different initial  $\sigma(0)$ under the same perturbation $(0,0)$ to $x$ and $y$.  Parameters: $a=b=1$, $ \omega = 5$, $c_1=-0.9, c_2=-0.7, c_3=0.5$, and $\varepsilon=0.01$.}
	\label{conger_speeds}
 \vspace{-15pt}
\end{figure}

\begin{example}
    In Fig.~\ref{conger_speeds}, we consider two initial values for $\sigma$: $\sigma(0) = -0.6$ and $\sigma(0) = -0.2$. Perturbations are introduced by setting $x(0) = y(0) = 0.1$ at $t=0$. Comparing the left and right plots, we see that as $\sigma$ approaches the bifurcation threshold at $\sigma = 0$, the recovery time for the perturbed neural activity variables $x$ and $y$ increases significantly, indicating a slower return to the equilibrium. \TiaQED
\end{example}

This example illustrates the phenomenon of critical slowing down as $\sigma$ drifts towards $\sigma =0$. The following theorem provides a theoretical foundation for this.

\begin{theorem}
	Let $\col(x(t),y(t),\sigma(t))$ be the solution of the system \eqref{main} that starts from $\col(x_0,y_0,\sigma_0)$, where $c_2<\sigma<0$, and $x_0,y_0$ satisfy
	\begin{equation}\label{lemma4_initial}
		x_0^2+y_0^2 < a-\sqrt{a^2+\sigma_0/b}.
	\end{equation}
	Let  $T> 0$ be the first time instant such that
	\begin{equation}\label{lemma4_final}
		x^2(T)+y^2(T) \ge a-\sqrt{a^2+\sigma(T)/b}.
	\end{equation}
	 Denote $r (t)\coloneq x^2(t) + y^2(t)$. Then, for any $0 \le t < T$, it holds that 
	 \begin{equation}\label{exponentially_rate}
	 	r (t) \le e^{-t \mu(t)} r (0),
	 \end{equation}
	 where 
	 \begin{equation}\label{def_mu}
	 	\mu(t) \coloneq -2\left(\sigma(t) +2ab r(0)-br^2(0) \right). 
	 \end{equation}
\end{theorem}

\begin{proof}
	The time derivative of $r(t)$ can be derived as 
	\begin{align*}
		\dot r (t ) = 2r(t)h(r(t),\sigma(t))
	\end{align*}
	with $h(r,\sigma)\coloneq \sigma +2ab r-br^2$. By the assumptions in \eqref{lemma4_initial} and \eqref{lemma4_final}, it holds that
	\begin{equation*}
		\dot r(t)<0, \forall t \in[0,T).
	\end{equation*}
	Consequently, $r(t) < r(0)<  a-\sqrt{a^2+\sigma_0/b}$. Since, given a fixed $\sigma$, $g(r,\sigma)$ is increasing in $r$ for $r\in [0, a-\sqrt{a^2+\sigma_0/b}]$, it holds that 	
	\begin{align*}
		\dot r (t ) \le  2r(t)h(r(0),\sigma(t))=-\mu (t) r(t). 
	\end{align*}
	Solving this inequality by taking $\mu(t)$ as a constant, one obtains the inequality \eqref{exponentially_rate}. 
\end{proof}

	Given any $t'\in[0,T)$, $\mu(t')$ represents the slowest convergence rate over the time interval $t\in[0,t']$. For small values of $t'$, $\mu(t')$ can be approximated by 
	\begin{equation}\label{def_mu0}
		\mu(0) = -2\left(\sigma(0) +2ab r(0)-br^2(0) \right),
	\end{equation}
	as $\sigma(t)$ does not change significantly. The reason is attributed to the slow timescale characterized by $\varepsilon\ll 1$, as evident from the definition of $\mu(t)$ in \eqref{def_mu}.  The convergence rate $\mu(0)$ highly depends on $\sigma(0)$, i.e., the instantaneous value of $\sigma$ at the moment when the perturbation is introduced. From \eqref{def_mu0}, $\mu(0)$ decreases as $\sigma(0)<0$ increases for fixed $r(0)$. 
	When $\sigma(0)\to 0$, it can be derived that $\mu(0)\to 0$. 
	
	These observations provide rigorous theoretical foundations for critical slowing down. Existing studies usually derive recovery rates for critical slowing down by linearization (e.g., see \cite{scheffer2009early, acharya2024predictive}). This approach is valid only when perturbations are sufficiently small, ensuring that the system remains close to its equilibrium. In contrast, our method constructs the convergence rate for any perturbation, as long as it does not drive the system beyond its region of attraction.

    Determining the system's recovery rate enables the detection of early warning signs for seizures, especially when there is a noticeable slowdown in recovery speed. From \eqref{def_mu0}, the instantaneous convergence rate can be calculated if $\sigma(0)$ and $r(0)$ are known. While $r(0)$ can be computed by measuring the neural activity variables  $x$ and $y$, the excitability  $\sigma(0)$ is typically not directly measurable. However, $\sigma(0)$ can be approximated indirectly from the observed values of $x$ and $y$.  Using the approximation $\mu(0)\approx \mu(t)$ and applying inequality \eqref{exponentially_rate}, it follows that
	\begin{equation}\label{estm_paramet}
		\sigma(0) \approx \frac{1}{2t} \ln \left(\frac{r(t)}{r(0)}\right) -2ab r(0)+br^2(0).
	\end{equation}


This analysis provides a way to develop detection strategies for identifying early warning signs of upcoming pathological oscillations. Next, we propose a detection algorithm.  



\subsection{Detection algorithm}
\begin{figure}[t]
	\centering
	\includegraphics[scale=1]{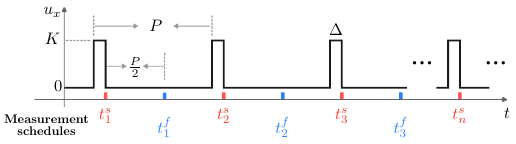}
	\vspace{-20pt}
	\caption{Schematic of periodically perturbing neural activity by stimulation-like pulses and measuring the system's responses to them.}
	\label{inputs_for_detection}
 \vspace{-15pt}
\end{figure}

As shown in Algorithm~\ref{control_placement}, we periodically perturb and sample the states of the system \eqref{stochastic}. 
As illustrated in Fig.~\ref{inputs_for_detection}, at time instants $t_1, t_2, \dots, t_n, \dots$, which satisfy $t_{n+1}-t_n = P$ for any $n$, control inputs are injected into the system. These inputs consist of pulses with width $\Delta$, mimicking electrical signals used in brain stimulation. After each pulse, the states $x$ and $y$ are measured according to a predesigned schedule:
\begin{align*}
	t_n^s \coloneq t_n+\Delta, \hspace{0.5cm}
	t_n^f \coloneq t_n^s+\frac{P}{2},  \hspace{0.5cm}	n=1, 2, \dots.
\end{align*}
Then, we compute $r^s_n = x^2(t_n^s )+y^2(t_n^s )$ and $r^f_n = x^2(t_n^f )+y^2(t_n^f )$.
Subsequently, we estimate the value of $\sigma(t)$ at $t=t_n^s$ following \eqref{estm_paramet}, that is,
\begin{equation}\label{estm_paramet_n}
	\sigma_n = \frac{1}{P} \ln \left(\frac{r^f_n }{r(0)}\right) -2abr^s_n +b(r^s_n )^2.
\end{equation}
we set a threshold $\bar \sigma$, slightly less than the critical point $\sigma=0$. When the estimated value $\sigma_n$ exceeds $\bar \sigma$, it indicates that a critical transition is imminent. This event serves as an early warning marker for impending seizures. 

\begin{algorithm}[t]
	\caption{Critical slowing down detection}
	\label{control_placement}
	\begin{algorithmic}[1]
		\State \textbf{Input:}  Discrete time instants $\{t_n: n=1,2, \dots\}$, $\{t^s_n: n=1,2, \dots\}$, and $\{t^f_n: n=1,2, \dots\}$, $P, K, \Delta>0,\bar \sigma$
		\State \textbf{Initialize:} $n=1,\sigma_0=-a^2b$ 
		\For{$t\ge 0$}
		\If{$t\in[t_n,t_n^s]$}
		\State $\zeta_x(t)=\zeta_y(t)=K$
		\Else
		\State $\zeta_x(t)=\zeta_y(t)=0$
		\EndIf
		\If{$t= t_n^s$ or $t= t_n^f$}
		\State Measure $x(t)$ and $y(t)$\\
		\hspace{30pt}If $t= t_n^f$: compute $\sigma_{n}$ by \eqref{estm_paramet_n}, and  $n=n+1$\\
		\hspace{26pt} If $\sigma_{n}>\bar \sigma$: \underbar{early biomarker detected}
		\EndIf
		\EndFor
	\end{algorithmic}
\end{algorithm}

\begin{figure}[t]
	\centering
	\includegraphics[scale=0.75]{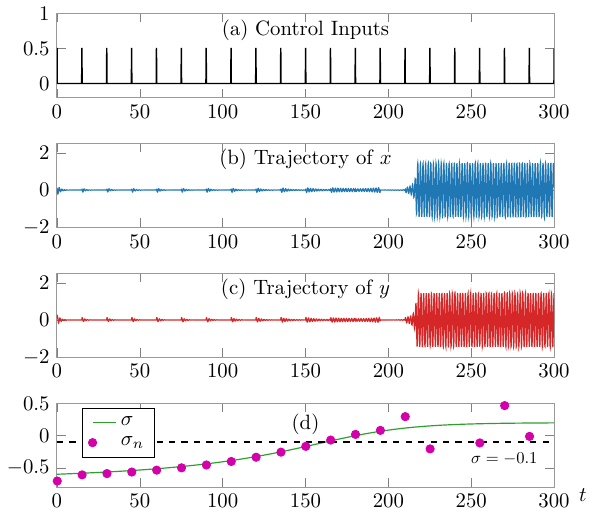}
	\vspace{-10pt}
	\caption{Detection of critical points. (a) Periodic injection of impulses into the system. (b)-(c) The inputs induce perturbations in the system's states, $x$ and $y$. (d) Following Algorithm~\ref{control_placement}, $\sigma$ is estimated to identify early warning signs of critical transitions. A threshold of $\bar \sigma=-0.1$ is set, values beyond this threshold indicate the detection of an event. Parameters: $a=b=1, \omega=4, c_1 =-0.9, c_2=-0.7, c_3=0.2$, and $\varepsilon=0.1$.}
	\label{dectation_example}
 \vspace{-15pt}
\end{figure}

\begin{example}
	To demonstrate the proposed algorithm, we simulate a system in Fig.~\ref{dectation_example}. Initially, the system operates near the steady state, reflecting normal neural activity. However, with the initial parameter $\sigma$ set above $c_2$, the system will eventually transition from stable normal activity to pathological oscillations.	
	To predict such critical transitions in advance, we configure the parameters in Algorithm~\ref{control_placement} as follows: $P = 15$, $K = 0.5$, and $\Delta = 0.2$. The pulses applied to the system are shown in Fig.~\ref{dectation_example}-(a), where the control impulses periodically perturb the system around its equilibrium. By measuring the states $x$ and $y$, we estimate and monitor the excitability parameter $\sigma$ using \eqref{estm_paramet_n}.
	
	A threshold $\bar{\sigma} = -0.1$ is used to signal impending critical transitions when $\sigma_n > \bar{\sigma}$. As shown in Fig.~\ref{dectation_example}-(d), the method successfully estimates the true $\sigma$ when it is below zero, predicting critical transitions effectively. Note that the estimation of $\sigma$ is no longer accurate after the transition takes place, because it is designed to predict approaching transitions. Further study is needed to develop approaches to estimate $\sigma$ after the critical transition since it might be helpful in designing adaptive control schemes to prevent seizures. 
\end{example}

\subsection{Event-based feedback control via precursor detection}
In this section, we demonstrate how detecting precursors that precede seizure-like oscillations, characterized by critical slowing down, can be leveraged to regulate epileptic dynamics, preventing ictal transitions to pathological oscillations. Specifically, we introduce linear feedback control to regulate the system, leading to the controlled dynamics of $x$ and $y$ expressed as
\begin{align*}
	\dot x &= - \omega y + x f(x, y, \sigma)+\zeta_x(t)+u_x(t), \\
	\dot y &= \hphantom{-}\omega x + y f(x, y,\sigma)+\zeta_y(t)+u_y(t),
\end{align*}
where $
	u_x=-F(t)x, u_y=-F(t)y. 
$
The feedback control gain is event-based, depending on whether a precursor event has been detected:
\begin{equation*}
	F(t)=\begin{cases}
		0, \text{ for } t^f_n<t \le t^f_{n+1}, \text{ if } \sigma_n \le \bar \sigma\\
		F, \text{ for all } t> t^f_n, \text{ otherwise.} 
	\end{cases}
\end{equation*}

\begin{figure}[t]
	\centering
	\includegraphics[scale=0.75]{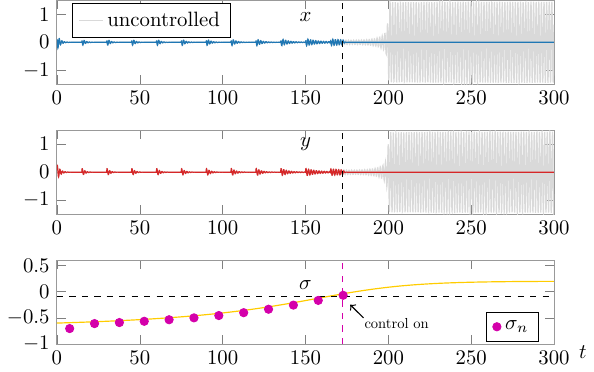}
	\vspace{-10pt}
	\caption{Illustration of event-based feedback control depending on detection of critical slowing down events. Parameters are the same as those in Fig.~\ref{dectation_example}.}
	\label{control}
 \vspace{-15pt}
\end{figure}

As shown in Fig.~\ref{control}, the feedback control with $F=1.4$ is activated upon detection of a pre-defined event. Compared to the uncontrolled scenario (gray trajectories), this control scheme prevents the system from transitioning into pathological oscillations. For simplicity, active probing is halted after detecting an early warning signal ($\zeta_x(t)$and $\zeta_y(t)$ set to 0). Future work will focus on designing adaptive and optimized control strategies that estimate the excitability parameter in real time by continuing to probe and measure the system.

\section{Discussion}

In this paper, we introduce a slow-fast system to model epilepsy dynamics, where the slow variable represents gradual changes in excitability. We analyze its multistability, derive regions of attraction for stable states, and examine the recovery speed after perturbations, establishing a theoretical foundation for critical slowing down. Based on these findings, we propose an algorithm that uses active probing of neural activity to detect critical slowing down events as precursors to seizures. We also present a proof-of-concept event-based control system to prevent pathological oscillations.
In the future, it is of interest to extend our work to stochastic models which capture the noisy nature of real-world brain dynamics.  Additionally, we aim to extend our approach beyond single neuronal populations to understand network-level epileptiform activity.

\bibliographystyle{IEEEtran}
\bibliography{\references}

\end{document}